\documentclass[proof]{WileyASNA-v1}

\articletype{Article Type}%

\received{26 April 2018}
\revised{6 June 2018}
\accepted{6 June 2018}

\begin{document}

\title{The Fermi GBM GRBs' multivariate statistics}

\author[1,2,3]{Istvan I. Racz*}

\author[1,3]{Lajos G. Bal\'azs}

\author[1,3]{Zsolt Bagoly}

\author[2]{Istvan Horvath}

\author[1,3]{Viktor T\'oth L.}

\authormark{RACZ \textsc{et al}}

\address[1]{\orgdiv{Department of Physics of Complex Systems}, \orgname{E\"otv\"os Lor\'and University}, \orgaddress{\state{Budapest}, \country{Hungary}}}

\address[2]{\orgdiv{Department of Natural Science}, \orgname{National Univesity of Public Services}, \orgaddress{\state{Budapest}, \country{Hungary}}}

\address[3]{\orgdiv{Konkoly Observatory, Research Centre for Astronomy and Earth Sciences}, \orgname{Hungarian Academy of Sciences}, \orgaddress{\state{Budapest}, \country{Hungary}}}

\corres{*\email{racz@complex.elte.hu}}

\presentaddress{5-th floor, office: 5.80, 1/A P\'azm\'any P\'eter s\'et\'any, 1117 Budapest, Hungary}

\abstract{Studying the GRBs' gamma-ray spectra may reveal some physical information of Gamma-ray Bursts. The Fermi satellite observed more than two thousand GRBs. The FERMIGBRST catalog contains GRB parameters (peak energy, spectral indices, intensity) estimated for both the total emission (fluence), and the emission during the interval of the peak flux. 
We found a relationship with linear discriminant analysis between the spectral categories and the model independent physical data. 
We compared the Swift and Fermi spectral types. We found a connection between the Fermi fluence spectra and the Swift spectra but the result of the peak flux spectra can be disputable. We found that those GRBs which were observed by both Swift and Fermi can similarly discriminate as the complete Fermi sample. We concluded that the common observation probably did not mean any trace of selection effects in the spectral behavior of GRBs.
}


\jnlcitation{\cname{%
\author{Istvan I. Racz}, 
\author{Lajos G: Balazs}, 
\author{Zsolt Bagoly}, 
\author{Istvan Horvath}, and 
\author{Viktor Toth L.}} (\cyear{2018}), 
\ctitle{The Fermi GBM GRBs' multivariate statistics}, \cjournal{AN}, \cvol{2018;XX:X--X}.}


\maketitle


\section{Introduction}

Gamma-ray bursts (GRBs) are the most energetic transients in the Universe \citep{Kumar}, and their spectral energy distribution (SED) represents basic information on the physical processes responsible for the observed characteristics of these objects \citep{Kumar6}. 
If we study the gamma energy range of the SED, it can be well approximated by a combination of a small number of power law functions \citep{grb_func, Kumar5}, which show that the nonthermal (synchrotron, inverse Compton) processes play probably a fundamental role in the radiation \citep{meszaros_fireball,prompt_emission,peter_meszaros_2015}. Sometimes we can find trace of the presence of a thermal component in the SED \citep{thermal_components,Ryde01,Ryde02}.

The Fermi Gamma-ray Space Telescope (FGST, Fermi satellite) provides the scientific community a wealth of high quality data on several astrophysical objects. The Fermi was launched in 2008 into a Low Earth orbit of $\approx$565 km and it has two instruments: the all-sky Gamma-ray Burst Monitor (GBM) can detect GRBs in the 8 keV to 40 MeV range \citep{fermi_gbm1,fermi_gbm2, fermi2} and the Large Area Telescope (LAT) can observe the 20 MeV to 300 GeV energy range photons \citep{fermi_lat_technical,fermi_lat}.

The Neil Gehrels Swift Observatory (formally known as 'Swift' Space Telescope) \citep{bat,xrt,uvot} has three different instruments: the Burst Alert Telescope (BAT), the X-ray Telescope (XRT), and the UV/Optical Telescope (UVOT). The instruments can detect in the energy ranges of 15--150 keV, 0.3--10 keV, and 170--600nm for the BAT, XRT, and UVOT, respectively. In contrary to GBM the BAT only have 1.4 sr field of view, but the position accuracy is a magnitude better than as of the Fermi GBM (a few arcminutes vs some degrees).

In general we can use the Band function \citep{band, Kumar7} to describe the observed SED within the limits of the statistical inference, but several GRBs can be well approximated by simpler functional forms (power law, cutted power law, Comptonized) \citep{comptonized1,smothly,band_smoothly,comptonized2}. Several attempts were made in fitting more complex spectra to the observed gamma energy distributions, but in most cases they were not really successful \citep{thermal1,thermal3,thermal2}. However, the Band function at the bright long GRBs' prompt radiation is often inadequate to fit the data \citep{peter_meszaros_2014B}. It is worth mentioning that these analytic models do not really have a physical background, although, they can be used to interpret some physical parameters (e.g. low- and high-energy spectral index or peak energy).

The single power law has two free parameters: the $\alpha$ spectral index and the $A$ amplitude. To normalize the model to the energy range under consideration a pivot energy (${\mbox{E}}_{\mbox{piv}}$) can be defined. It was fixed at 100 keV for GRBs detected by GBM.

The Comptonized model is an exponentially cutoff power law, a subset of the Band function in the limit when $\beta \rightarrow \infty$. It has three free parameters: the $\alpha$ low-energy spectral index, the $A$ amplitude and $\mbox{E}_{\mbox{peak}}$. Similarly to the power-law model ${\mbox{E}}_{\mbox{piv}}$ is again fixed at 100 keV.

The smoothly broken power law spectra is characterized by one breakpoint with flexible curvature and is able to fit spectra with sharp or smooth transitions between the low- and high-energy power laws. This model has five parameters and was introduced by \citet{smothly}.

Finally, Band's GRB function is usually considered as the standard for fitting GRB spectra. This function has four free parameters: $\alpha$ and $\beta$ the low- and high energy spectral indices, respectively, the $A$ amplitude, and the $\mbox{E}_{\mbox{peak}}$ energy.

In our recent work \citep{racz2018} we revealed an ordering of the spectra into a power law -- Comptonized -- smoothly broken power law -- Band series with contingency and correspondence analysis. 

There are some model independent quantities such as the duration, fluence and flux bear important physical information, all these can be obtained from analyzing the observed gamma radiation \citep{Kumar2,Kumar3}. In the following we try to find relationships between the spectral data and model independent quantities \citep{peter_meszaros_2014}.

The GBM has recorded a break in the majority of the GRBs' spectral energy distribution \citep{fermi1,fermi3}. This break depends on the photon flux quantity, therefore we cannot fit faint sources well with functions that differ from the power law.

In this work we studied the relationship between the GRBs' model-independent parameters and spectral properties. We compared the Swift and Fermi observations and analyzed the correspondence of the best fitted models.

\section{GBM Spectral Data}
\label{speda}

There is a big catalog from the GBM observations which were classified as GRBs. There are two types of spectra in the Fermi data: one referring to the peak flux (designated with pflx) and the other one to the fluence (designated with flnc). The peak flux spectra relates to the photons collected in the time bin around the intensity peak of the prompt emission (the time bin equals to 1024 ms at the bursts of $T90 > 2 s$ duration and 64 ms if $T90 < 2 s$). Both in the pflx and flnc spectra four different  model spectra were fitted to every burst event and the best fitting model was indicated by a likelihood-based statistic. The four fitted model spectra are the power law, the Comptonized, the Band and the smoothly broken power law.

The FERMIGRBST\footnote{\url{https://heasarc.gsfc.nasa.gov/W3Browse/fermi/fermigbrst.html}} catalog \citep{fermi4, fermi5} contains several physical parameters derived from the rough data delivered by the Fermi GBM detectors. The database used in our paper contains more than 2200 records (number of GRBs). One set of the data (e.g. duration, peak flux, fluence) is related to the burst in general and the other one were obtained from fitting the models, along with the best fitted one.

Only the best fitted models in both sets (pflx and flnc) of model spectra were taken into account in our analysis, ignoring the remaining three spectra and their estimated parameters. As for the model independent parameters we used the $T50$ and $T90$ durations, the 64/256/1024 ms flux in the 10--1000 keV energy range, the 64/256/1024 ms (BATSE standard) flux in the 50--300 keV energy range, the 10--1000 keV energy band fluence, and the 50--300 keV energy band BATSE fluence.

\section{Multivariate Analysis of the Data}
\label{mulan}

\subsection{Analysis of the contingency tables}
\label{ctanal}

There are N observations with several variables for each observation, and all events were classified into one of 'A' mutually exclusive categories and one of 'B' mutually exclusive categories. Then we can make the contingency table (an $A\times B$ matrix) by cross-tabulating the data. Unassociated row and column variables are referred to as independent. 

The definition for independence is that

\begin{equation}
P(row_i,column_j)=P(row_i)\cdot P(column_j)
\end{equation}

\noindent for all i,j, where $P$ is the probability. In our case pflx and flnc spectral categories can be considered as independent so we can test our null-hypothesis $(H_0)$ with the following chi-square test statistic:

\begin{equation}
\chi^2=\Sigma^{row}_{i=1} \Sigma^{column}_{j=1} \frac{(O_{ij}-E_{ij})^2}{E_{ij}}
\end{equation}

\noindent where $O_{ij}$ is the observed and $E_{ij}$ is the expected number of events (frequencies).

The usage of this method is beneficial in determining whether there is dependence, but as the strength of that association depends on the degrees of freedom as well as the value of the test statistic, it is not easy to interpret the strength of association.

The Pearson's residuals (which are also known as Pearson's contingency coefficient) is one method to provide an easy way to interpret the strength of association. Specifically, it is:

\begin{equation}
\textrm{Pearson's\ residuals}=\frac{(O_{ij}-E_{ij})}{\sqrt{E_{ij}}}
\end{equation}

\noindent This residual have approximate Normal distribution with mean 0 (no association).

\subsection{Discriminant Analysis}

Discriminant analysis aims to recognize difference between groups in the multivariate parameter space, orders membership probabilities to the cases and one may use this scheme for classifying additional ones not having assigned group memberships. We use this technique to look for differences in the distributions of GRBs in the parameter space defined by the variables listed in the Fermi GBM catalog.

There are some different methods for the problem stated above. In this case we used linear discriminant analysis (LDA) originally proposed by \citet{LDA1}; see \citet{LDA2} for a review. Further mathematical background can be found in Sect.~3.3.1 of \citet{racz2018}.

LDA presents in all professional statistical software packages. We used the \textcolor{red}{\tt lda()} procedure available in the $MASS$ library \citep{MASS} of the R statistical environment.

\subsubsection{LDA of the GBM Data}
\label{lda}

We analyzed more than two thousand GRBs of the the Fermi GBM catalog with the LDA method. We used the the following model independent variables: $T90$, $T50$, fluence, batse fluence, 64ms/256ms/1024ms flux, batse 64ms/256ms/1024ms flux. In the analysis we take both types of spectral fitting into account where the categorical variables were the best fitted model of both peak flux and fluence.

The LDA resulted in three discriminant functions (LD1,~LD2,~LD3) based on the parameters above. The mean values of the discriminant functions within the different spectral types and significance obtained. The significance of the discrimination between the spectral types was shown by the output of this procedure. The Wilks' lambda parameters can be calculated based on the probability, which is the level of significance.

We may use the complete sample without taking the different uncertainties of the data given in the catalog into consideration, but the Fermi group published a better sample. \citet{fermi1}, however, defines a subsample of 'GOOD' having uncertainties below a certain limit. For our study we used LDA with the data fulfilling these criteria. These criteria do not differentiate between the classical GRB groups (short, intermediate, and long) \citep{hoi2002, ripa}.

\begin{figure}
\centering
\includegraphics[width=3.9cm]{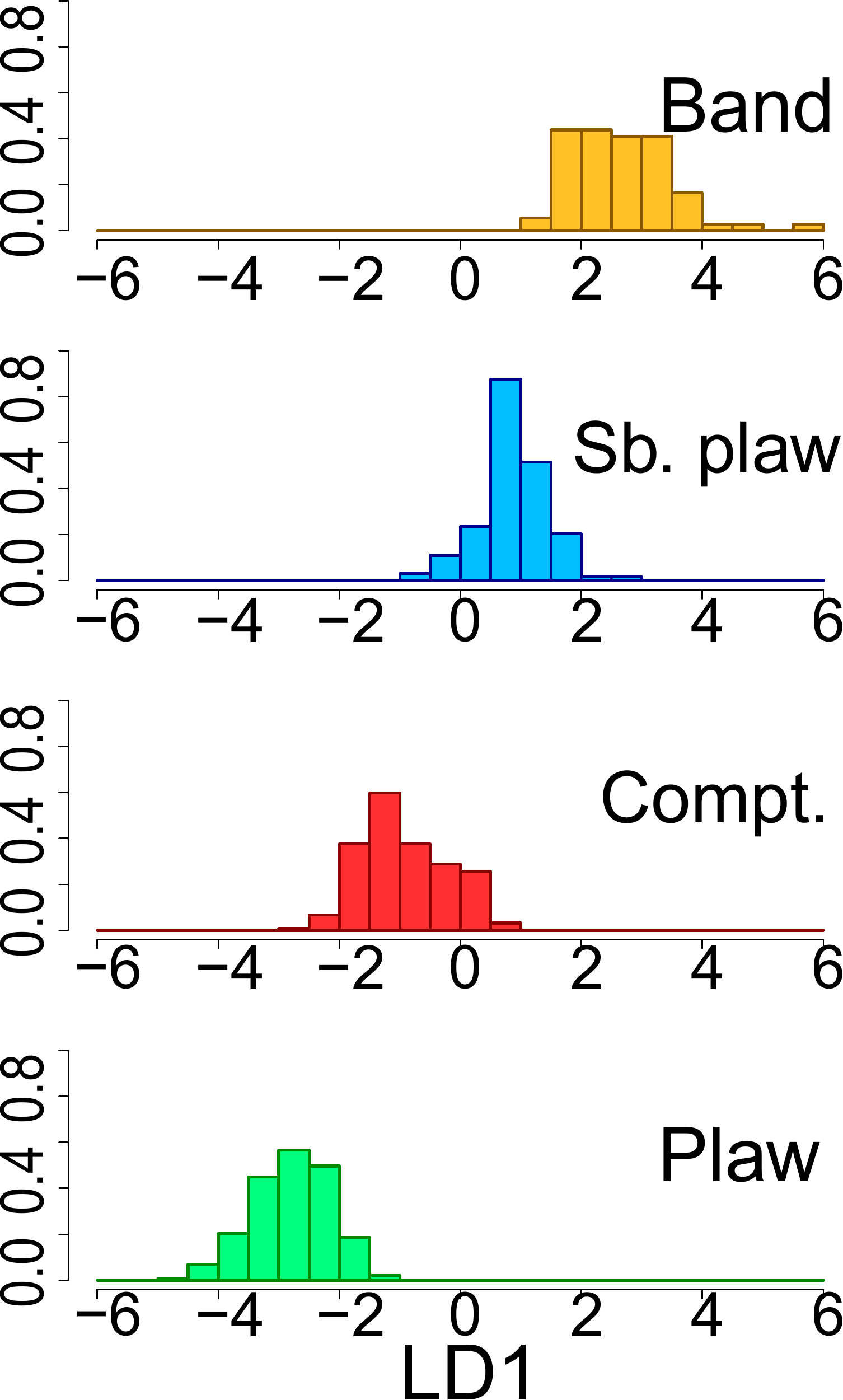}
\includegraphics[width=3.9cm]{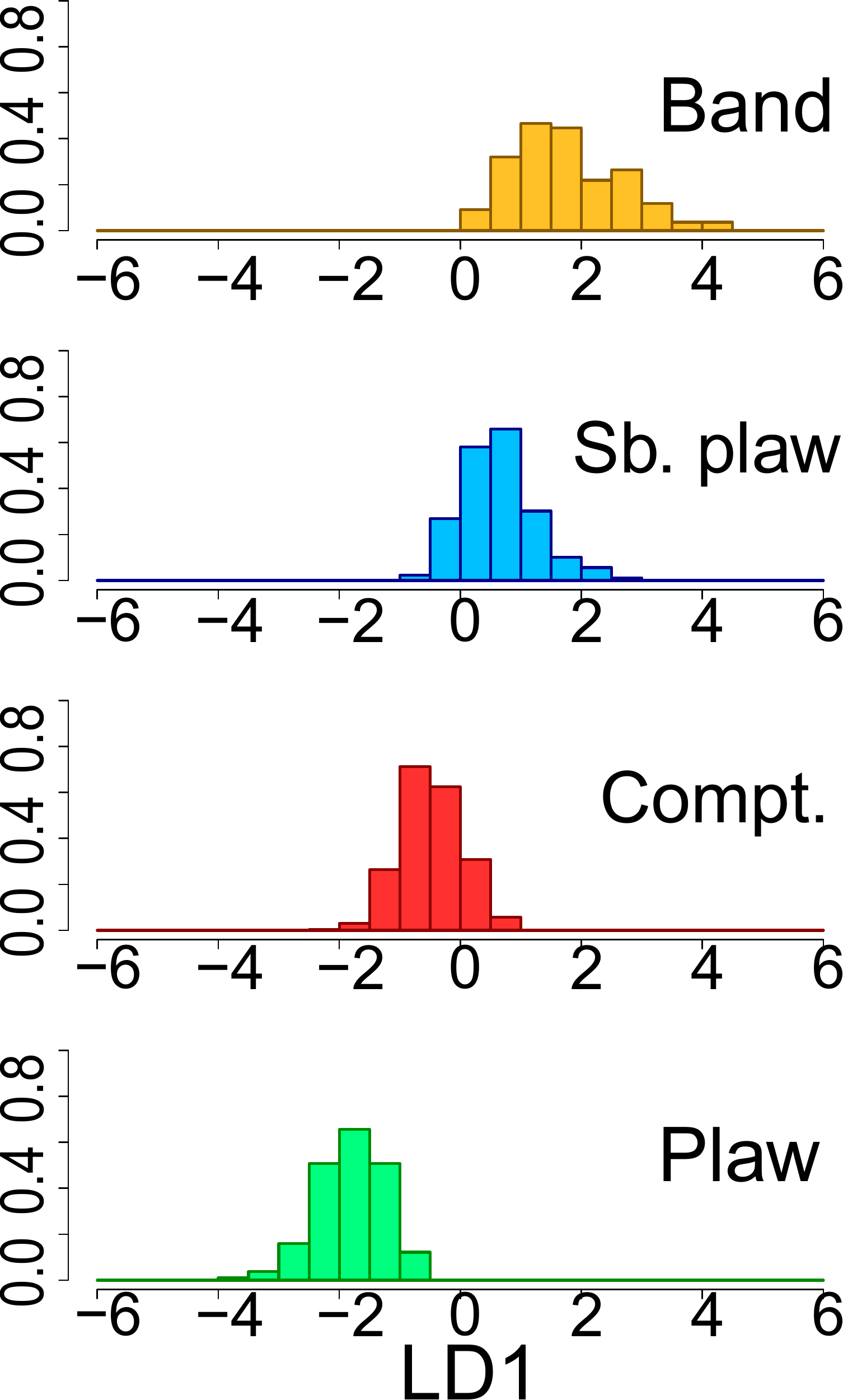}
\caption{\label{fig:lda_full} The 'GOOD' sample GRBs' distribution of the first discriminant function between the spectral classes and types. On the left side the 'peak flux' types and on the right the 'fluence' type models. We found that the difference between the classes are bigger in the 'peak flux' models but both (\textit{pflx} and \textit{flnc}) significances are very high.}
\end{figure}

Our results obtained from the peak flux and fluence spectra are also shown in Fig.~\ref{fig:lda_full}. One can see that in both cases the first linear discriminant function (LD1) separates the different spectral types well. 
The results show that fitting the model spectra on the peak flux is much more efficient for the significance values if we compare the two plots. The Band and the smoothly broken power law merged into each other, but the LDA could separate the power law and the Comptonized types well (in both of the classification types).

The greatest discriminated distances appear to be between the Band and the power law classes, with the remaining two classes being between them \citep{racz2018}. The fluence spectra contain much more photons, thus more informations about the GRBs even though in this work we found the surprising and significant effect that the level of discrimination is higher in case of the peak flux than the fluence spectra.

Performing LDA in this case would result in improved significances, and as we obtained the  LD2 of the peak flux spectra became significant as well (sign. level: $<10^{-3}$). 
The LDA also show that the duration of the GRB is not a significant discriminator, thus our results are valid for all groups of GRBs.

\section{Associating the Swift-Fermi GRBs}

Both the Swift and the Fermi satellites detected several thousand events from which more than a thousand were triggered GRBs. 
It can be declared boldly that the two satellites opened a new area in studying GRBs. There are significant differences in the covered energy range because the Fermi is sensitive from a few keV to a few GeV, in contrast the Swift can observe the radiation up to 150 keV. 
We can find several GRBs which both Fermi and Swift had detected.

The NASA Swift GRB Table \footnote{\url{https://swift.gsfc.nasa.gov/archive/grb_table/}} was made using the GRB observations and contains the trigger time and position data for more than 1100 Swift discovered GRBs until the May of 2018.

\begin{figure}
\includegraphics[width=0.48\textwidth]{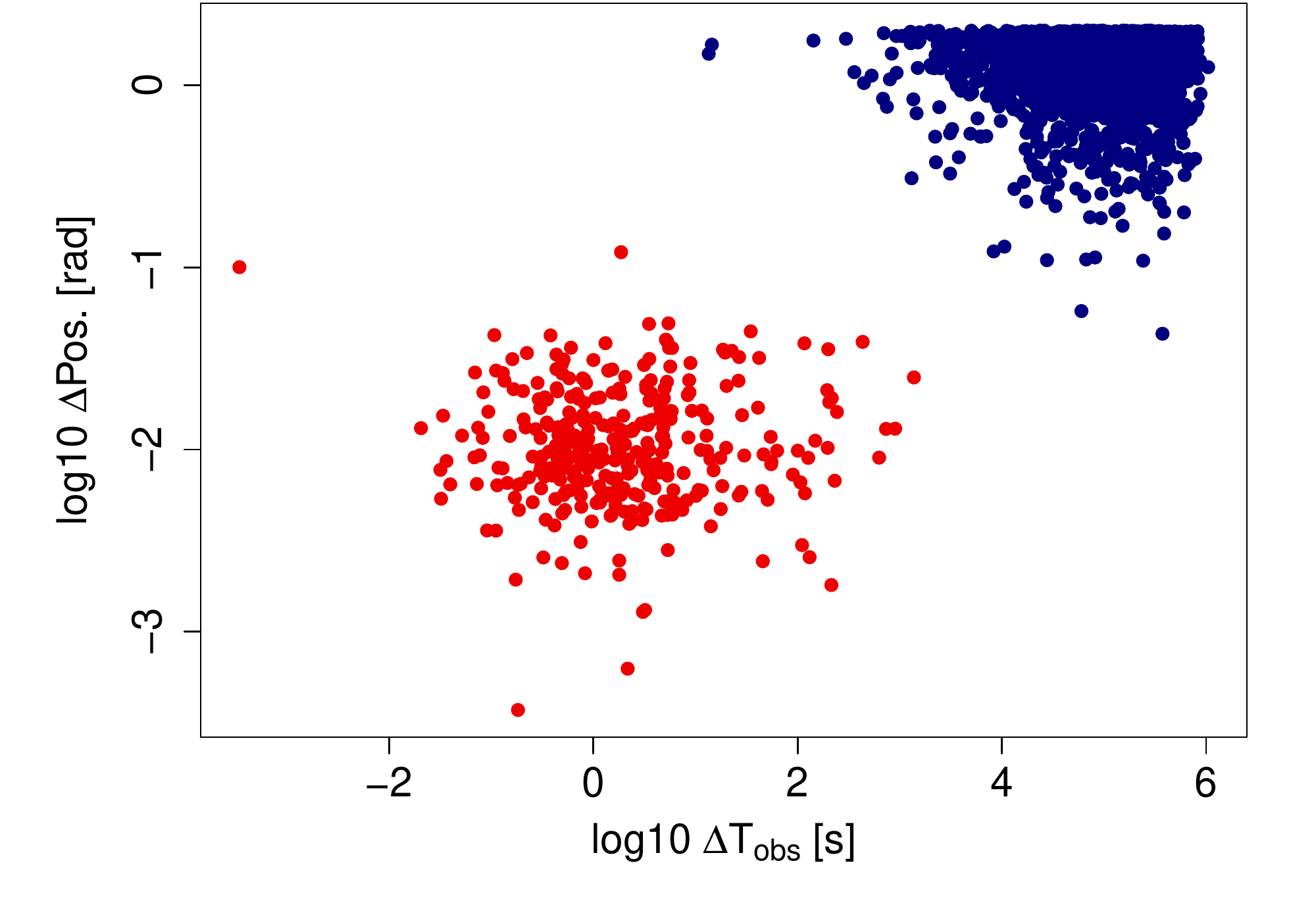}
\caption{Matching the Swift-Fermi GRBs. To find the Fermi-Swift data pairs of the shortest time and spatial differences nearest neighbor procedures were used. Then Model-Based Clustering was used to select the Swift-Fermi pairs. We found a well separated group (red dots) in the time/position plane where the observation time difference is smaller than 30 minutes and the coordinates are within 10 degrees. \label{fig: fermi_swift}}
\end{figure}

We compared the objects presented in the GBM catalog with those recorded by Swift to get the simultaneously detected GRBs. We looked for the closest neighbor in between the two catalogs in position and time to find the matching objects.

We used the \textcolor{red}{\tt knn()} procedure from the $class$ library of the R environment \citep{MASS} to assign every Fermi GRBs a Swift counterpart with the shortest time and spatial differences. With this procedure we got a 2D plane from the difference of observation times and positions where the points on this plane show the closest associations between the Fermi-Swift events. The results can be seen in Fig.~\ref{fig: fermi_swift}. 

The Model-Based Clustering procedure (\textcolor{red}{\tt Mclust()}) from the $mclust$ package \citep{mclust1,mclust2} provided a well separated group, which can also be seen by the naked eye. This group of GRBs is clearly defined by the time difference being less than 30 minutes and the coordinate difference being less than 10 degrees. 

{Finally, we checked the accidental multiple associations and omitted these 8 events. In our previous work we presented 292 Fermi-Swift pairs from 2069 Fermi and 1122 Swift GRBs, now, after 1 year, we found 316 pairs from 2307 and 1205 GRBs. We note that the Swift GRB Table contains Fermi discovered and Swift observed GRBs, but in that table there are only 52 events.}

\subsection{Comparing the Fermi-Swift spectral classes}

\begin{figure}
\includegraphics[width=0.495\linewidth, trim= 0 0 0 0]{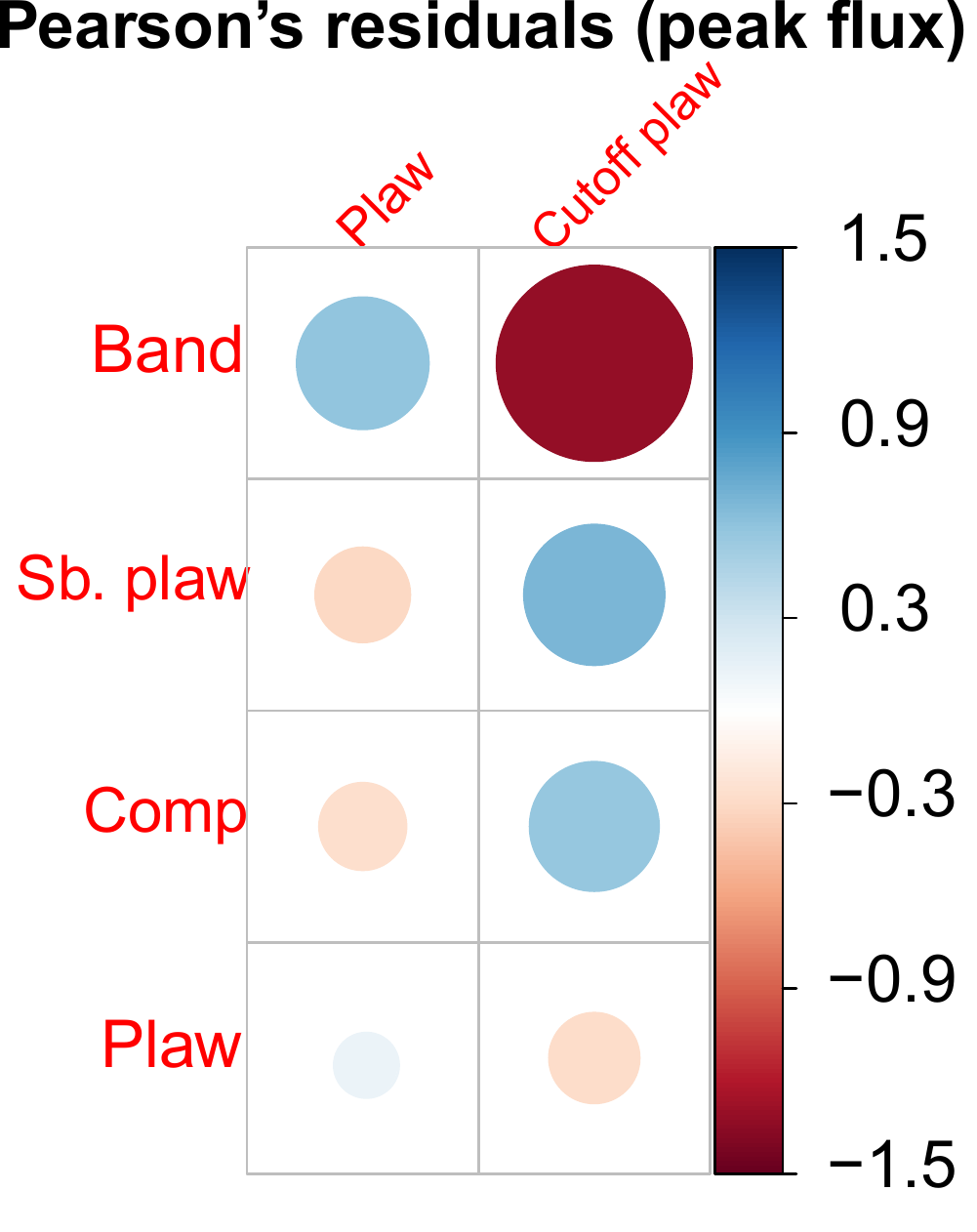}
\includegraphics[width=0.495\linewidth, trim= 0 0 0 0]{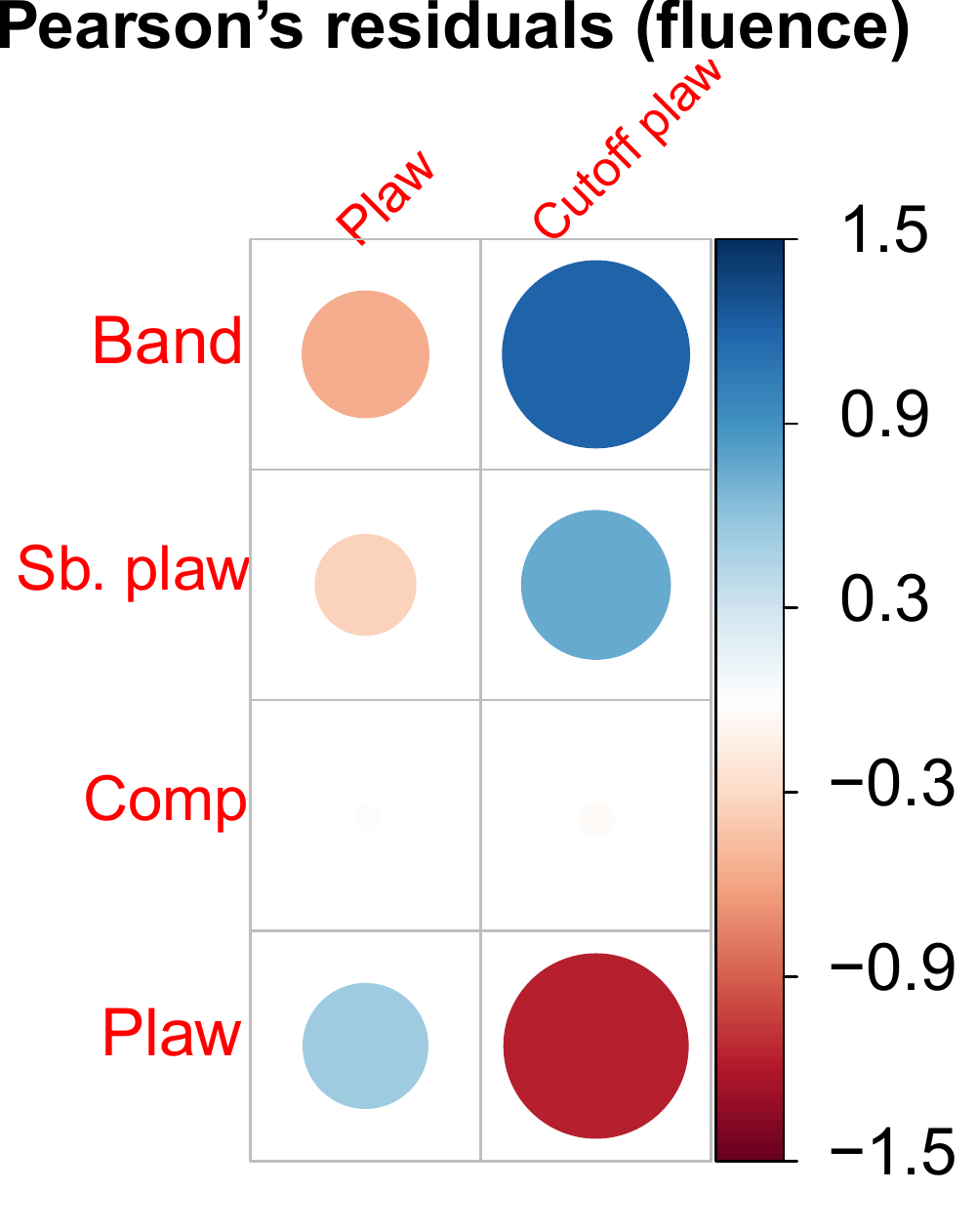}
\caption{The strength of relationship between the Swift and Fermi spectral classes. Left: the spectra during the peak flux, right: the spectra during the whole burst. The circle size and the color show the strength of the association. The blue color means positive and red means negative correlations. Flnc follows the expectations but at the Pflx type there is a remarkable deficit for the Band -- cutoff power law group. \label{pearson_fermi_swift}}
\end{figure}

In contrast to the Fermi, the Swift GRB Table contains only two spectral classes (power law and cutoff power law) and does not distinguish the spectral types (over the peak flux or during the duration), therefore we only have two category variables. We studied the relationship between the Fermi and the Swift spectral classes. The results of the contingency analysis -- the strength of association (Pearson's coefficient) -- is shown in Fig.~\ref{pearson_fermi_swift}. The Fermi fluence type spectral classes match well because we cannot see a power law -- non-power law (Comptonized, broken, or Band) association. Also, the cutoff power law connected the Band and smoothly broken power law. Contrarily the peak flux type spectra show a surprising Band--power law relationship. The reason of this strange association can probably be that the Swift classified the spectra over the entire burst duration similar to the Fermi fluence type and the spectrum changed in time. Several papers have found similar variability on the spectrum (among others \citep{meszaros_2006,Kumar,racz2018}). The spectral evolution is best shown by the Band spectra, because more GRBs have the Band spectral class over the peak flux time.

\subsection{LDA of common GRBs}

\begin{figure}
\centering
\includegraphics[width=3.9cm]{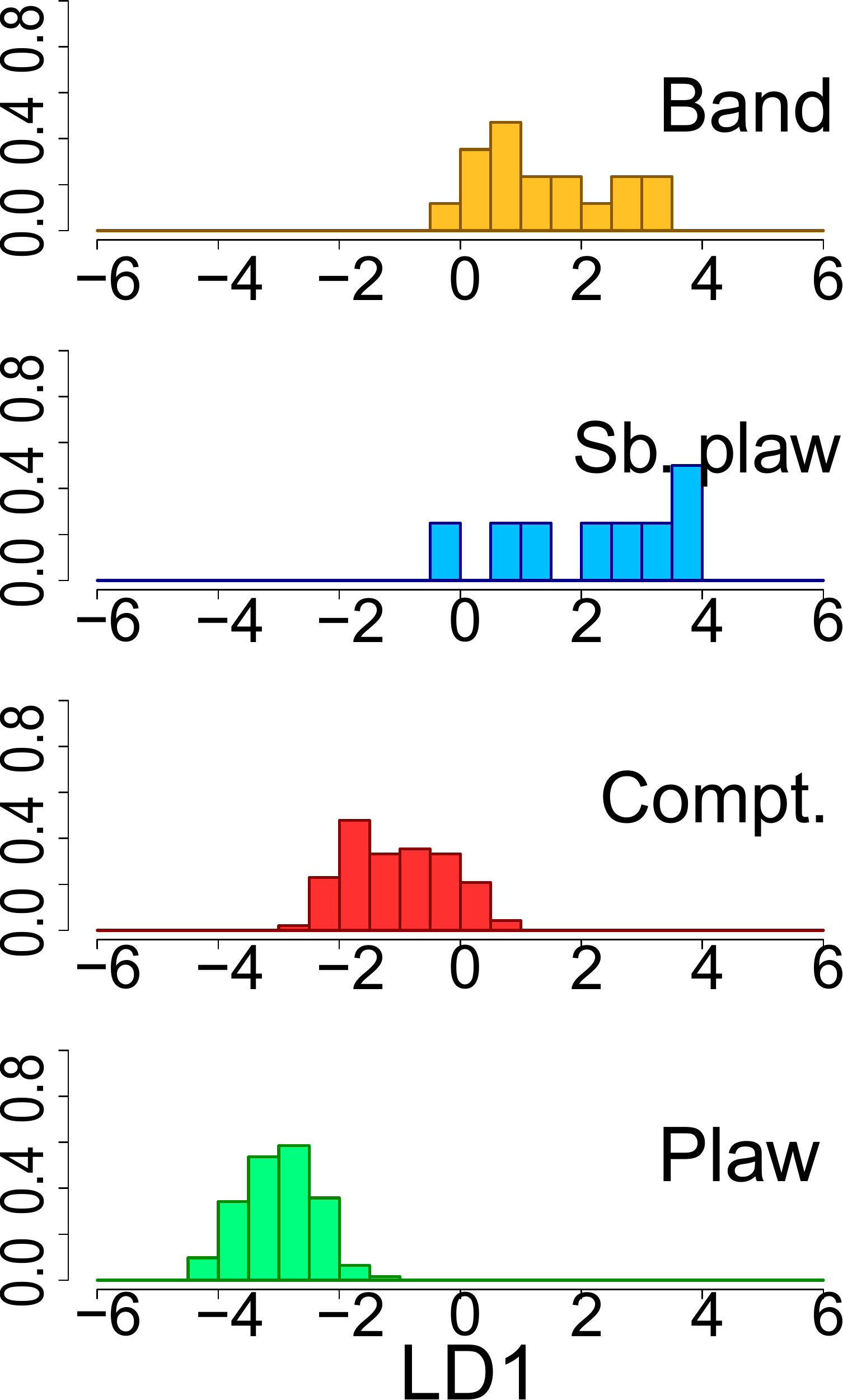}
\includegraphics[width=3.9cm]{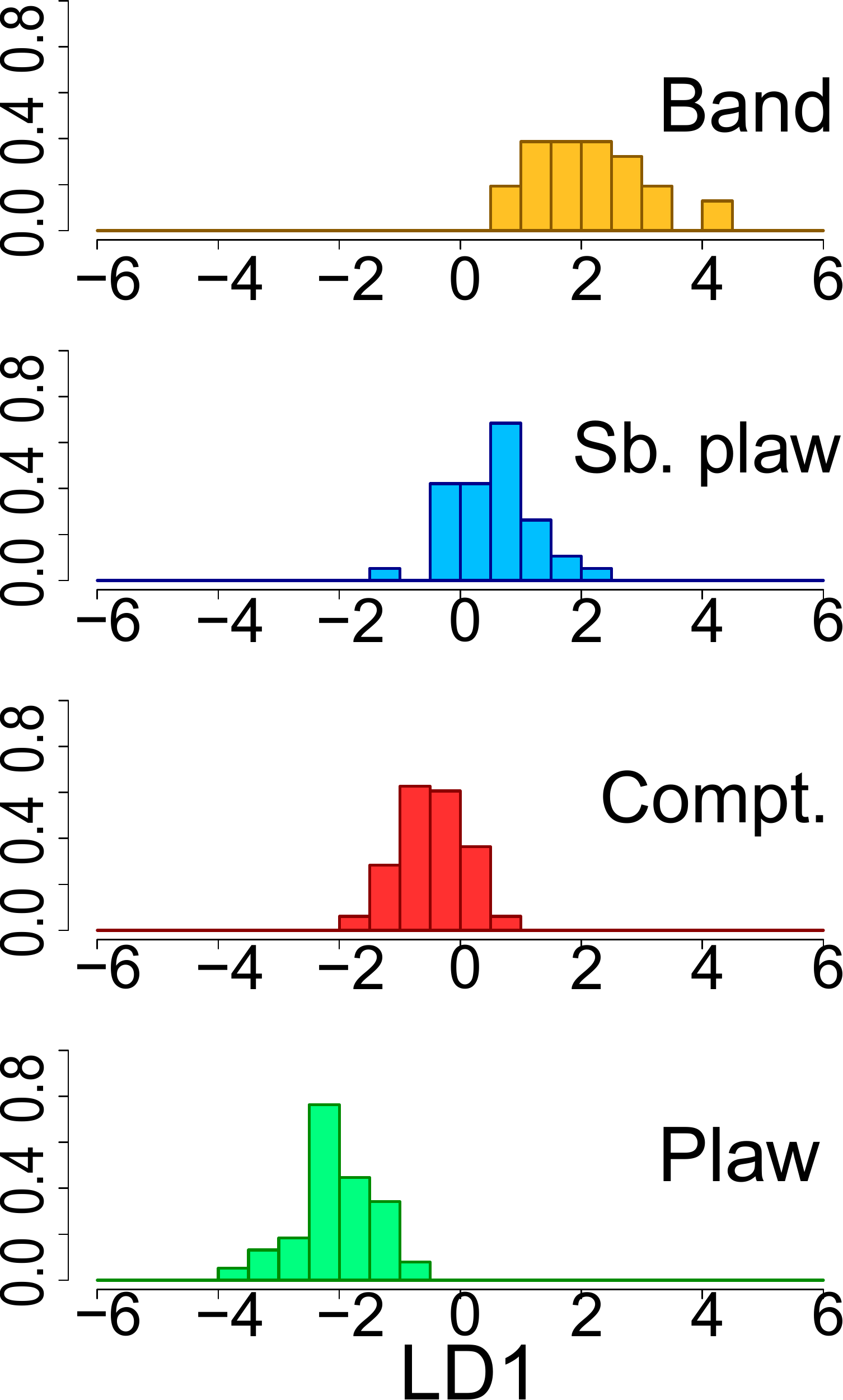}
\caption{The first discriminate function of those GRBs which have been observed by both Fermi and Swift. We can see the Fermi peak flux type spectra on the left, and the fluence type spectra on the right. It is similar to the original 'GOOD' sample in spite of the fact that the Band and smoothly broken power law were merged on the peak flux type.\label{Fig: lda_common}}
\end{figure}

By using LDA we have examined the GRBs which have been observed by both Swift and Fermi at the same time. Similarly to Sect.~\ref{lda}., the first linear discriminate function can be seen in Fig.~\ref{Fig: lda_common}. It shows consistency with the previous results from the 'GOOD' sample. The Band and the smoothly broken power law spectral classes are merged on the peak flux spectra type, but the peak flux still appears to be a better discriminator. The primary cause of the significant overlap might be the low number of objects, because there are only a few GRBs that both satellites were able to detect, and their best fitted spectral class was the smoothly broken power law. We were unable to find any evidence that the GRBs' simultaneous observation would cause a selection effect in the spectral classes.

\section{Summary}

Gamma-ray bursts, which are the brightest transient in the Universe, can be detected -- besides the gamma-ray energy range -- in almost all wavelengths. Generally the spectral energy distribution can be well-approximated by a small number of power law functions. The Fermi Gamma-ray Space Telescope discovered more than 2000 GRB events, and 2 types of spectra were made from the GRBs:  over the time range of the peak flux of the burst (peak flux type, pflx) and over the entire burst duration (fluence type, flnc). 4 spectral classes could be fitted on both types: Band, smoothly broken power law, Comptonized and power law. 

The best fitted spectral classes can be seen as category variables, and we are able to study the relationship between the model independent parameters (duration, total energy, peak flux) and these categories. We used the procedure of linear discriminant analysis and we have found a significant separation between the spectral classes on both spectral types. Our results also show that the peak flux spectra can discriminate better, even though the fluence spectra contain much more photons, thus more informations about the GRBs.

We paired Fermi and Swift observations, and we studied the association between the Fermi and Swift spectral classes. The Swift spectra were made over the entire duration of the burst like the Fermi fluence type spectra, these classes were well-matched. In the case of 'pflx', we have found the trace of spectral evolution because there is a deficit on the Band -- cutoff power law associations.

The LDA method was used to analyze the GRBs which have been detected by both the Swift and the Fermi satellites to test whether the two satellites are observing different populations of GRBs. Our results are similar to the previous ones: we were unable to find any significant evidence of any selection effect. The small difference can be explained with the small number of objects in our sample.
terpretation of data as well as the preparation of the manuscript.

\section*{Acknowledgments}

Supported by the \fundingAgency{Hungarian OTKA} \fundingNumber{NN-111016} grant and supported by the New National Excellence Program of the \fundingAgency{Hungarian Ministry of Human Capacities} \fundingNumber{UNKP-17-3}. We are grateful to Rebeka Bogner and Agnes J. Hortobagyi for their help in preparing the manuscript. We are thankful to the Anonymous Referee for his helpful suggestions concerning the presentation of this paper.

\subsection*{Author contributions}
All authors participated in acquisition, analysis and interpretation of data as well as the preparation of the manuscript.

\subsection*{Financial disclosure}

None reported.

\subsection*{Conflict of interest}

The authors declare no potential conflict of interests.

\bibliography{racz.bib}

\section*{Author Biography}
\begin{biography}{\includegraphics[width=60pt,height=70pt]{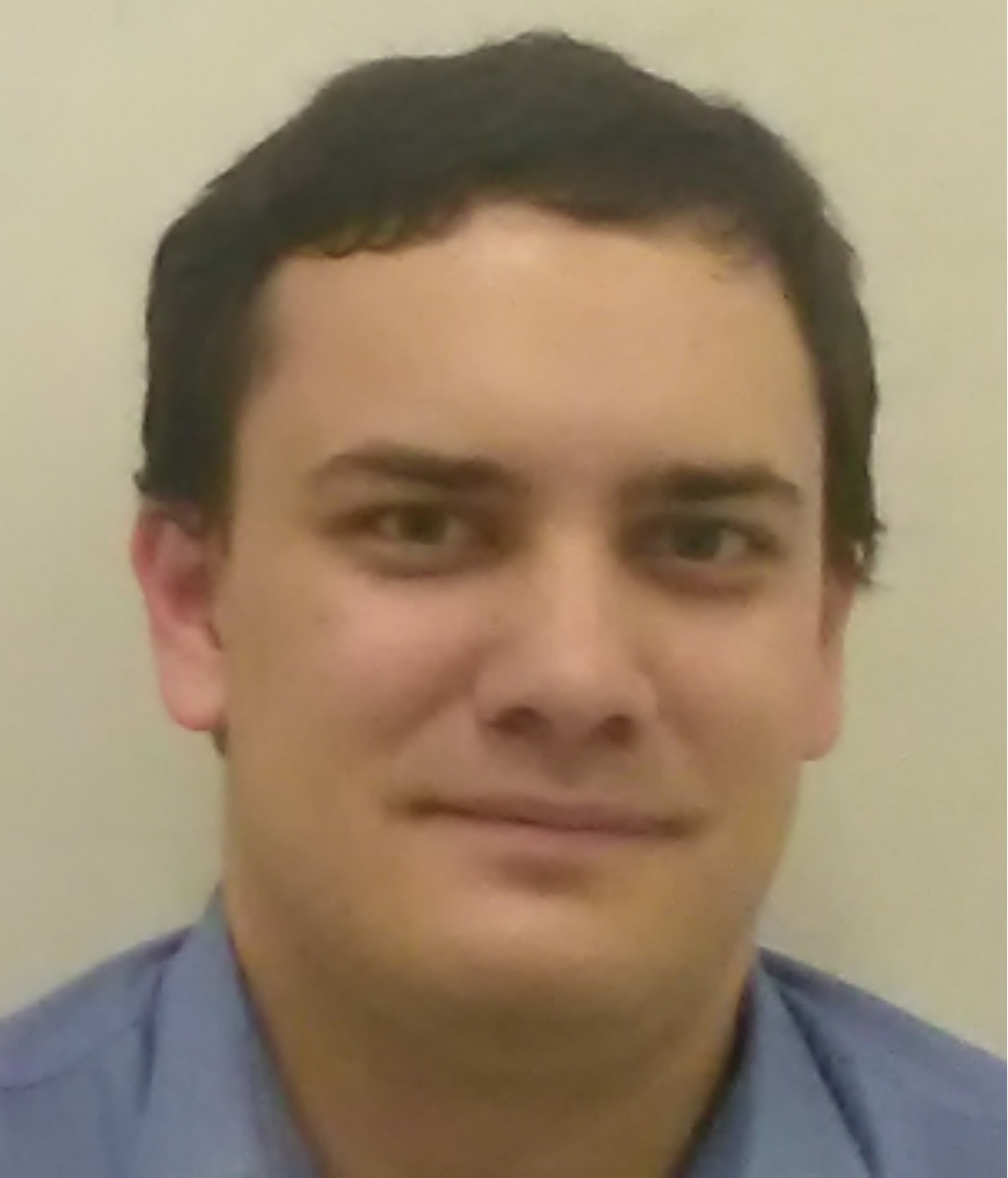}}
{\textbf{Istvan I. Racz.} 
PhD Student, E\"otv\"os Lor\'and University, Budapest, Hungary; Assistant lecturer, National Univesity of Public Services, Budapest, Hungary
Education: PhD School of Physics (Particle Physics and Astronomy), 2015 --, E\"otv\"os Lor\'and University, project: Gamma ray bursts and their Cosmic environment.
MSc in Astronomy, 2015, E\"otv\"os Lor\'and University, thesis: Starformation in Planck cold clumps, supervisor: L. Viktor T\'oth. BSc in Physics, 2012, E\"otv\"os Lor\'and University, thesis: Expansion of the Universe, supervisor: Istvan Csabai.

Research interests: Statistical analysis of the spatial distribution of Gamma-ray bursts, studying the central engine and GRB afterglow with X-ray and Gamma spectral fitting.

}
\end{biography}

\end{document}